\documentclass[aps,prl,twocolumn,showpacs,groupedaddress]{revtex4}

\usepackage[]{graphicx, pslatex}
\usepackage{epstopdf}

\begin{document}


\title{Tuning the Resonance in High Temperature Superconducting Terahertz Metamaterials }


\author{Hou-Tong Chen}
\email{chenht@lanl.gov}
\author{Hao Yang}
\author{Ranjan Singh}
\author{John F. O'Hara}
\author{Abul K. Azad}
\author{Stuart A. Trugman}
\author{Q. X. Jia}
\author{Antoinette J. Taylor}
\affiliation{MPA-CINT, MS K771, Los Alamos National Laboratory, Los Alamos, New Mexico 87545}

\date{\today}

\begin{abstract}
In this Letter we present resonance properties in terahertz metamaterials consisting of a split-ring resonator array made from high temperature superconducting films. By varying the temperature, we observed efficient metamaterial resonance switching and frequency tuning with some features not revealed before. The results were well reproduced by numerical simulations of metamaterial resonance using the experimentally measured complex conductivity of the superconducting film. We developed a theoretical model that explains the tuning features, which takes into account the resistive resonance damping and additional split-ring inductance contributed from both the real and imaginary parts of the temperature-dependent complex conductivity. The theoretical model further predicted more efficient resonance switching and frequency shifting in metamaterials consisting of a thinner superconducting split-ring resonator array, which were also verified in experiments. 
\end{abstract}


\maketitle

Metamaterials consisting of metallic elements have enabled a structurally scalable electrical and/or magnetic resonant response, from which exotic electromagnetic phenomena absent in natural materials have been observed~\cite{Smith2000}. Metals provide high conductivity that is necessary to realize strong electrical/magnetic metamaterial response~\cite{Pendry1996,Pendry1999}. Metals, however, play a negligible role in active/dynamical metamaterial resonance switching and/or frequency tuning, which has been typically accomplished through the integration of metamaterials with other natural materials (e.g. semiconductors) or devices, and by the application of external stimuli~\cite{Shadrivov2006,Padilla2006,Chen2006,Kim2007,Chen2008,Chen2009,Driscoll2009,Tao2009,Dani2009}. It is essentially the modification of the metamaterial embedded environment that contributes to such previously observed functionalities.

Recently, there has been increasing interest in superconducting metamaterials towards loss reduction~\cite{Salehi2005,Wang2006,Ricci2005,Ricci2006,Ricci2007,Fedotov2010,Gu2010,Jin2010}. Significant Joule losses have often prevented resonant metal metamaterials from achieving proposed applications, particularly in the optical frequency range. At low temperatures, superconducting materials possess superior conductivity than metals at frequencies up to terahertz (THz), and therefore it is expected that superconducting metamaterials will have a lower loss than metal metamaterials. More interestingly, superconductors exhibit tunable complex conductivity over a wide range of values, through variation of temperature and application of photoexcitation, electrical currents and magnetic fields. Therefore, we would expect correspondingly tunable metamaterials, which originate from the superconducting materials composing the metamaterial, in contrast to tuning the metamaterial environment.

In fact, superconducting metamaterials have enabled diamagnetic response at very low frequencies, which may enable screening of static magnetic fields~\cite{Wood2007,Magnus2008}. In the microwave frequency range ($\sim$10~GHz), left-handed superconducting transmission lines have been introduced and their tunability has been realized with electrical currents or temperature~\cite{Salehi2005,Wang2006}. Negative index metamaterials comprised of niobium wires and split-ring resonators exhibited red-shifting in resonance frequency ($\sim0.6\%$) when the temperature was increased approaching the transition temperature $T_{\rm c}$~\cite{Ricci2006}. Further experimental work has demonstrated tunability through application of external dc or rf magnetic fields~\cite{Ricci2007,Jin2010}. At THz frequencies, low temperature superconductors may be not suitable for metamaterial applications, since with a smaller superconducting gap, Cooper pairs may be excited and broken by THz photons, and therefore high temperature superconductors (HTS) should be employed. In this Letter, we present THz metamaterials based on electric split-ring resonators (SRRs) made from epitaxial YBa$_2$Cu$_3$O$_{7-\delta}$ HTS films. These metameterials exhibit temperature-dependent resonance strength and frequency, which reveal some interesting tuning features not previously observed. Finite-element numerical simulations and theoretical modeling are performed to understand the underlying tuning mechanism. 

\begin{figure}[h!]
\includegraphics[width=3in]{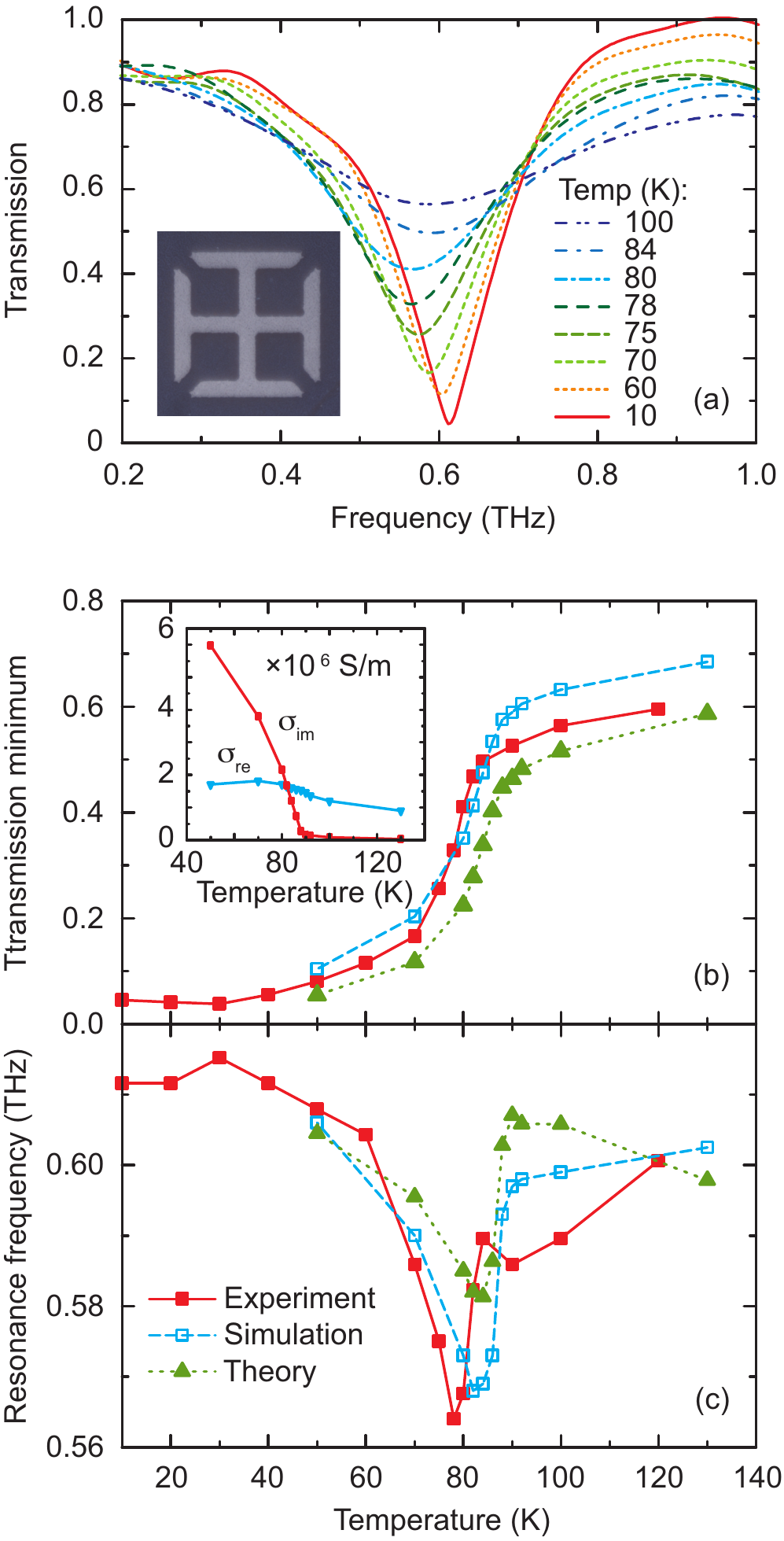}%
\caption{\label{Fig1} (color online). (a) THz Transmission amplitude spectra of the 180~nm thick YBCO metamaterial at various temperatures. (b) Transmission minimum and (c) corresponding resonance frequency as functions of temperature, from experiments, numerical simulations, and theoretical calculations. Inset to (a) illustrates a microscopic image of a single YBCO SRR, where the lighter colored area is YBCO. Inset to (b) shows the real and imaginary parts of the complex conductivity at 0.6~THz of an unpatterned 180~nm thick YBCO film. } 
\end{figure}

The epitaxial YBa$_2$Cu$_3$O$_{7-\delta}$ (YBCO) films with $\delta = 0.05$ were prepared using pulsed laser deposition on 500 $\mu$m thick (100) LaAlO$_3$ (LAO) substrates. The transition temperature was measured to be $T_{\rm c} = 90$~K. Square arrays of electric SRRs, with the unit cell shown in the inset to Fig.~\ref{Fig1}(a), were fabricated using conventional photolithographic methods and wet chemical etching of the YBCO films. The YBCO SRRs have a thickness of $d = 180$~nm or 50~nm, outer dimensions of $l = 36$~$\mu$m, line width of $w = 4$~$\mu$m and gap size of $g = 4$~$\mu$m, and the arrays have the periodicity of $p = 46$~$\mu$m. Terahertz time-domain spectroscopy (THz-TDS) incorporated with a continuous flow liquid helium cryostat was used to characterize the YBCO films and metamaterials. Under normal incidence, the THz transmission spectra were measured as a function of temperature, using an LAO substrate as the reference. 

We focus our attention on the metamaterial fundamental resonance (the so-called \textit{LC} resonance) resulting from the circulating currents excited by the incident THz electric field~\cite{Chen2007}. In Fig.~\ref{Fig1}(a) we show the THz transmission amplitude spectra for the 180~nm thick YBCO metamaterial sample at various temperatures. At temperatures far below $T_{\rm c}$, e.g.~10~K, the metamaterial exhibits the strongest resonance, as indicated by the sharp THz transmission dip with a minimal transmission amplitude of 0.045 at 0.613~THz. This strong resonance is almost the same as in a metamaterial sample where the YBCO SRRs was replaced by gold SRRs with the same thickness and at the same temperature.  As the temperature increases, the resonance strength decreases, as seen by the broadening and reduction in amplitude of the transmission dip. The resonance frequency experiences a red-shifting, which reaches the lowest value of 0.564~THz near 80~K, resulting in a frequency tuning of 8\%. As the temperature further increases, the resonance strength continues to decrease, but the resonance frequency, on the other hand, shifts back to higher frequencies. The temperature-dependent transmission minimum and the corresponding resonance frequency are plotted in Figs.~\ref{Fig1}(b) and \ref{Fig1}(c), respectively. The results show that, at temperatures near 80~K, the transition of resonance strength is fastest and the resonance frequency exhibits a dip, not observed in previous work~\cite{Fedotov2010,Ricci2006,Gu2010,Jin2010}. We can exclude the LAO substrate as contributing to the metamaterial resonance tuning, because the features in the temperature-dependent resonance in the YBCO metamaterial (see Fig.~\ref{Fig1}) were not observed in a metamaterial sample where the YBCO was replaced by gold. In that gold SRR metamaterial sample, the resonance frequency shifting was imperceptible, and the resonance strength only slightly decreased with increasing temperatures. Additionally, through THz-TDS measurements, it turns out that the dielectric constant of the LAO substrate only exhibits a weak dependence on the temperature. Therefore, it is the temperature-dependent properties of YBCO film that are responsible for the observed metamaterial resonance tuning.

The complex conductivity of YBCO film can be expressed using the well-known two-fluid model~\cite{Tinkham1996}:
\begin{equation}
\tilde{\sigma}(\omega, T) = \frac{n e^2}{m^{\ast}} \left[ \frac{f_{\rm n}(T)}{\tau^{-1} - i \omega} + i \frac{f_{\rm s}(T)}{\omega} \right], \label{TwoFluid}
\end{equation}
where $f_{\rm n}$ and $f_{\rm s}$ are fractions of normal (quasiparticle) and superconducting (superfluid) carriers, respectively, with $f_{\rm n} + f_{\rm s} = 1$, $n$ is the carrier density,  $m^{\ast}$ is the carrier effective mass, and $\tau$ is the quasiparticle relaxation time. The real and imaginary parts of the complex conductivity are then:
\begin{eqnarray}
\sigma_{\rm re} & = & \frac{n e^2}{m^{\ast}} \frac{f_{\rm n}(T) \tau}{1 + \omega^2 \tau^2}, \label{Real}\\
\sigma_{\rm im} & = & \frac{n e^2}{m^{\ast}} \left[ \frac{f_{\rm n}(T) \omega \tau^2}{1 + \omega^2 \tau^2} + \frac{f_{\rm s}(T)}{\omega} \right]. \label{Imag}
\end{eqnarray}

Using THz-TDS we experimentally measured the conductivity of an unpatterned 180~nm thick YBCO plain film. The resultant real and imaginary parts of the complex conductivity at 0.6 THz are plotted as functions of temperature in the inset to Fig.~\ref{Fig1}(b). The real conductivity [Eq.~(\ref{Real})], which derives from the Drude response of quasiparticles in Eq.~(\ref{TwoFluid}), slowly increases when temperature decreases across $T_{\rm c}$ to about 70~K. It starts to decrease below 70~K, but not significantly, over the temperatures we measured down to 50~K.  In this temperature range, $\omega \tau \ll 1$ at the resonance frequency ($\sim$ 0.6~THz), the decreasing $f_{\rm n}(T)$ may be compensated by the increasing quasiparticle scattering time $\tau$~\cite{Gao1996}. At temperatures above $T_{\rm c}$, the imaginary conductivity is derived from the Drude response and is very small, since $f_{\rm s}(T>T_{\rm c}) = 0$. As the temperature decreases below $T_{\rm c}$, the second term in Eq. (\ref{Imag}) from the superfluid Cooper pair state becomes non-zero and results in the rapidly increasing imaginary conductivity, exceeding the real conductivity below 80~K. Using these experimental values of the YBCO complex conductivity at 0.6~THz, the metamaterial resonant response was simulated using commercially available finite-element simulation codes from COMSOL Multiphysics. The simulated transmission minimum and the corresponding frequency are plotted as functions of temperature in Figs.~\ref{Fig1}(b) and \ref{Fig1}(c), respectively, reproducing the experimental results.

The measured real conductivity of the YBCO film reveals less than 20\% change over the temperature range from 60~K to 90~K, where the resonance strength experiences a fast change and the resonance shifts its frequency. This variation of real conductivity cannot solely cause the observed large metamaterial resonance switching and frequency tuning. Both the real and imaginary parts of the complex conductivity have to be considered for the metamaterial resonance. The imaginary conductivity, which is due to the superfluid carriers and causes no loss, becomes dominant at low temperatures, and it is responsible for the enhancement in resonance strength. 

The resonance frequency is determined by the effective capacitance, $C$, inductance, $L$, and resistance, $R$, in the SRRs~\cite{Purcell1985}:
\begin{eqnarray}
\omega_0^2 = \frac{1}{LC} - \frac{R^2}{4 L^2}.
\label{ResonanceFrequency}
\end{eqnarray}
It has been shown that the kinetic inductance, which represents the kinetic energy storage in free electrons in metals, or Cooper pairs in superconductors, plays an important role in determining the metamaterial resonance frequency~\cite{Zhou2005,Ricci2006}. This effect underpins the red-shifting of the resonance frequency in niobium metal superconducting metamaterials operating near 10 GHz as the temperature increases and approaches $T_{\rm c}$~\cite{Ricci2006}. However, the back blue-shifting, shown in Fig.~\ref{Fig1}(c) between $\sim$80~K and $T_{\rm c}$, was not observed in niobium, and the model proposed in that work would not explain this effect when only the superfluid  state (i.e. imaginary conductivity) was considered~\cite{Ricci2006}. 

Here we consider a more general situation that the SRRs are fabricated from a conducting film (YBCO film in our case) with a complex conductivity $\tilde{\sigma}$ and thickness $d$. Such an unpatterned plain film can be modeled as a lumped impedance in an equivalent transmission line. By equating the (multiple) reflections or transmissions from the film and the transmission line model, this complex surface impedance (or sheet impedance with units of $\Omega$/square) of the unpatterned film can be derived as: 
\begin{eqnarray}
\tilde{Z}_{\rm S} = R_{\rm S} - i X_{\rm S} = Z_0 \frac{n_3 + i \tilde{n}_2 \cot (\tilde{\beta}d)}{\tilde{n}_2^2-n_3^2}, \label{SurfaceImpedance}
\end{eqnarray}
where the tildes over the variables indicate complex values, $Z_0 = 377$~$\Omega$ is the vacuum intrinsic impedance, $n_3 = \sqrt{\epsilon_{\rm LAO}} = 4.8$ is the LAO substrate refractive index, $\tilde{n}_2 = \sqrt{i \tilde{\sigma}/\epsilon_0 \omega}$ is the complex refractive index of the film, and $\tilde{\beta} = \tilde{n}_2 \omega/c_0$ is the complex propagation constant where $c_0$ is the light velocity in vacuum. Both $\tilde{n}_2$ and $\tilde{\beta}$ can be calculated from the experimental complex conductivity near the metamaterial resonance frequency. When $n_3 \ll |\tilde{n}_2|$, which is valid in our case, Eq.~(\ref{SurfaceImpedance}) can be further simplified as:
\begin{eqnarray}
\tilde{Z}_{\rm S} = i \frac{Z_0}{\tilde{n}_2} \cot (\tilde{\beta}d). \label{SurfaceImpedanceSimplified}
\end{eqnarray}

\begin{figure}[b!]
\includegraphics[width=3in]{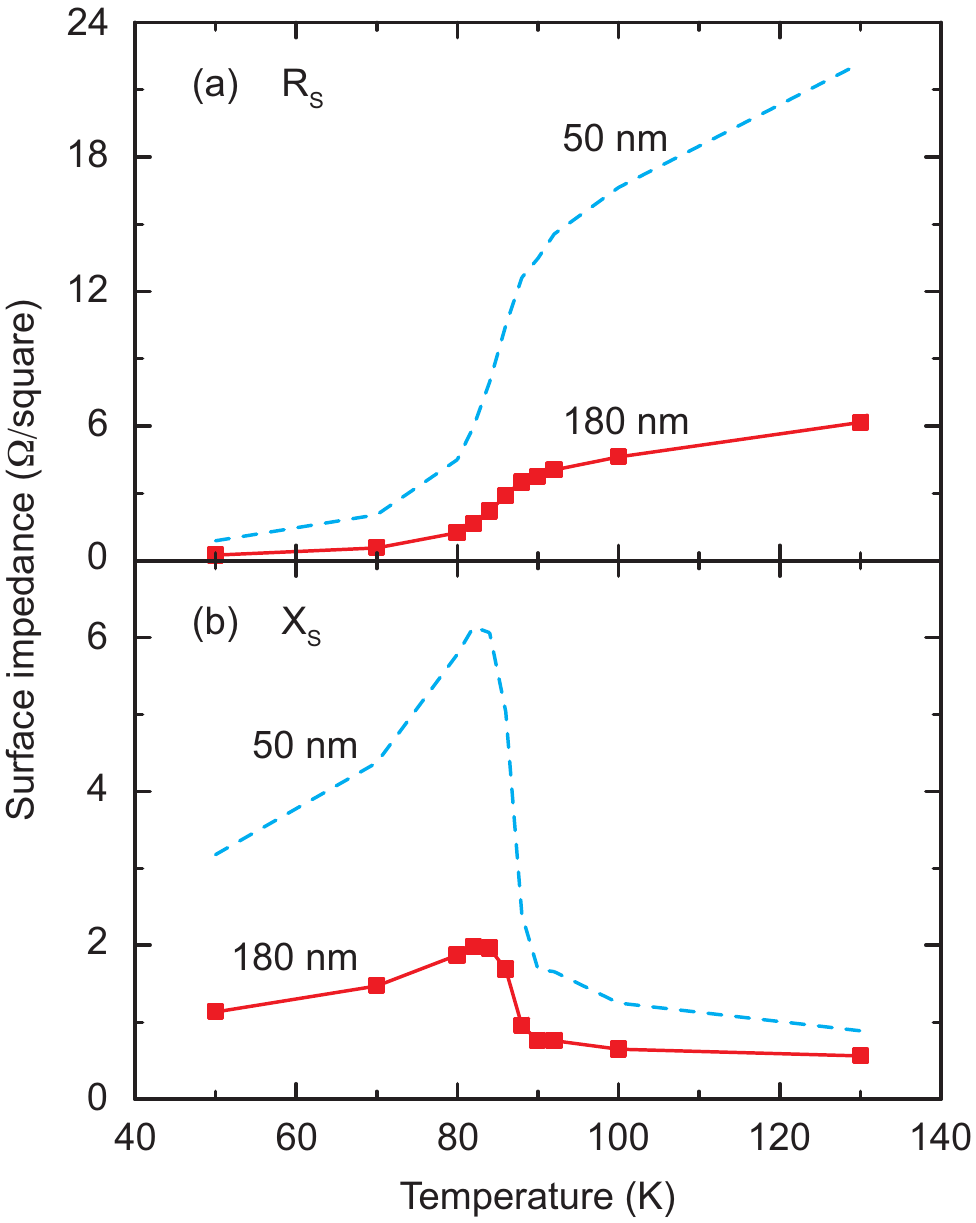}%
\caption{\label{Fig2} (color online). Complex surface impedance of the 180~nm thick unpatterned YBCO superconducting film calculated using the experimental complex conductivity at 0.6~THz. (a) Surface resistance $R_{\rm S}$, and (b) surface reactance $X_{\rm S}$. The dashed curves are for an assumed 50~nm thick YBCO film.} 
\end{figure}

From Eqs.~(\ref{SurfaceImpedance}) and (\ref{SurfaceImpedanceSimplified}), it is obvious that both the finite real and imaginary parts of the film refractive index $\tilde{n}_2$, and therefore the finite real and imaginary parts of the complex conductivity $\tilde{\sigma}$, contribute to the film surface resistance $R_{\rm S}$ and reactance $X_{\rm S}$. They are plotted in Fig.~\ref{Fig2} for the 180~nm thick YBCO film as functions of temperature, and are also calculated for a 50~nm thick YBCO film assuming its complex conductivity does not depend on the film thickness. 

The YBCO SRR array resistance $R$ (SRR reactance $X$ is zero at resonance) can be obtained by considering the nonuniform distribution of currents in a unit cell~\cite{Ulrich1967}: $R \cong [(A - g )/ w] R_{\rm S}$, where $A = 64$~$\mu$m is the median circumference of the (small) current loop. When temperature increases, the increasing SRR resistance $R$ accounts for the resonance damping and therefore the increasing transmission minimum. If we model the SRR array as a lumped resistor $R$ in the transmission line~\cite{OHara2007}, we can calculate the resonance transmission as a function of temperature, which is plotted in Fig.~\ref{Fig1}(b) and satisfyingly reproduces the experimental and simulated results. 


In order to correctly interpret the temperature-dependent resonance frequency shifting, additional inductance in SRRs has to be taken into account besides the geometric inductance $L_{\rm G}$. The geometric inductance represents the conventional inductance of the SRR loop and can be estimated~\cite{Terman1943} to be $L_{\rm G} \cong 4 \times 10^{-11}$~H. In contrast, the additional inductance $L_{\rm S}$ originates dominantly from the kinetic energy in superconducting carriers in the YBCO SRRs. This additional SRR inductance can be calculated using the above derived YBCO film surface reactance $X_{\rm S}$ and by considering the geometry and dimensions of the YBCO SRR: $L_{\rm S} \cong [(A - g )/ w] (X_{\rm S}/\omega)$. Therefore, the total SRR inductance becomes $L = L_{\rm G} + L_{\rm S}$. In order to obtain the metamaterial resonance frequency using Eq.~(\ref{ResonanceFrequency}), we estimate the SRR capacitance $C \cong 1.5 \times 10^{-15}$~F, from the above estimated geometric inductance $L_{\rm G}$ and the simulated resonance frequency $\omega_0 = 2 \pi \times 0.62$~THz assuming perfect conducting SRRs (i.e. $L_{\rm S} = 0$ and $R = 0$ ). The calculated temperature-dependent metamaterial resonance frequency is plotted in Fig.~\ref{Fig1}(c) along with the experimental and simulation results. Again, the theoretical result reproduces the frequency tuning features, though the overall tuning range is about half of the experimental and simulation data. 


The above calculations show that the temperature-dependent SRR resistance and additional inductance, due to the temperature-dependent complex conductivity of the YBCO film, play an important role in the resonance switching and frequency tuning. Eqs.~ (\ref{SurfaceImpedance}) and (\ref{SurfaceImpedanceSimplified}) further reveal that, for a fixed value of the real conductivity, which is approximately the case in our situation, the YBCO film surface reactance, and therefore the additional SRR inductance, reach the maximum value when the imaginary conductivity is approximately equal to the real conductivity, and vice versa. This is consistent with the experimental observations, where the metamaterial resonance frequency shifts to the lowest value when the real and imaginary parts of the YBCO complex conductivity cross each other. 

\begin{figure}[h!]
\includegraphics[width=3in]{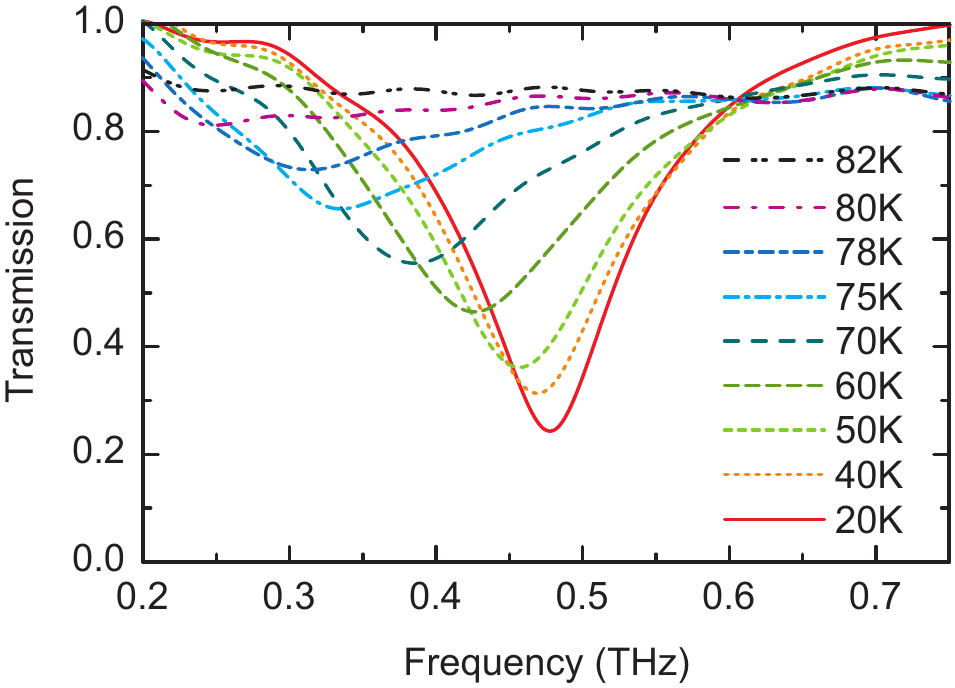}%
\caption{\label{Fig3} (color online). Temperature-dependent THz Transmission amplitude spectra of the 50~nm thick YBCO metamaterial.} 
\end{figure}

The results in Fig.~\ref{Fig2} suggest that metamaterials made from thinner YBCO superconducting films will have a lower resonance frequency, and will be more efficient in resonance switching and frequency tuning. In order to verify this prediction, we fabricated and characterized a second metamaterial sample from 50~nm thick YBCO film. The temperature-dependent transmission spectra are shown in Fig.~\ref{Fig3}. The resonance frequency at 20~K is measured to be 0.48~THz, which is significantly lower than that in the metamaterial sample from 180~nm thick YBCO film. When temperature increases, the resonance frequency continuously shifts to lower frequencies. It becomes 0.31~THz at 78~K, achieving a tuning range of 35\%. We did not observe the back shifting of resonance frequency due to the high resistance at temperatures above 80~K, which already completely damps the metamaterial resonance.

In conclusion, we have fabricated and characterized electric SRR-based metamaterials from high temperature superconducting YBCO films. We observed temperature induced metamaterial resonance switching and frequency tuning, which can be reproduced by finite-element numerical simulations using the experimentally measured complex conductivity of the YBCO film. We found that both the temperature-dependent real and imaginary parts of the complex conductivity of the superconducting film have to be consistently considered in order to achieve a correct interpretation. A theoretical model has been developed, taking into account the SRR resistance and additional inductance. Our modeling calculations were in good agreement with experimental observations and numerical simulations, and further predicted  more efficient resonance switching and frequency tuning with thinner YBCO metamaterials, which was also verified in experiments. We expect that such resonance tuning in superconducting metamaterials could also be realized dynamically through application of optical excitation, electrical currents, and/or magnetic fields. Although high temperature superconducting metamaterials may not be able to essentially address the loss issue at THz frequencies and beyond, they should enable the development of novel, multi-functional metamaterials. 

We acknowledge support from the Los Alamos National Laboratory LDRD Program. This work was performed, in part, at the Center for Integrated Nanotechnologies, a US Department of Energy, Office of Basic Energy Sciences Nanoscale Science Research Center operated jointly by Los Alamos and Sandia National Laboratories. Los Alamos National Laboratory, an affirmative action/equal opportunity employer, is operated by Los Alamos National Security, LLC, for the National Nuclear Security Administration of the US Department of Energy under contract DE-AC52-06NA25396.


\end{document}